\documentclass[fleqn,twoside]{article}
\usepackage{espcrc2}

\usepackage{graphicx}
\usepackage[figuresright]{rotating}

\newcommand{\AmS}{{\protect\the\textfont2
  A\kern-.1667em\lower.5ex\hbox{M}\kern-.125emS}}

\title{EDELWEISS-II : Status and Future}

\author{V. Sanglard\address[MCSD]{Institut de Physique Nucl\'eaire de Lyon, \\
	Universit\'e Claude Bernard Lyon 1, \\
	43, Bd du 11 Novembre 1918, \\ 
	69622 Villeurbanne Cedex, FRANCE}%
        \thanks{E-mail: sanglard@ipnl.in2p3.fr}, for the EDELWEISS collaboration}
       
\begin{document}

\begin{abstract}
%EDELWEISS is a direct search for WIMPs using cryogenic Ge ionization-phonon detectors and located in the Modane 
%Underground Laboratory. We summarize the final results of the EDELWEISS-I experiment obtained with up to nearly one kg of
%detectors. The increased exposure confirms previous results. We also report on the preparations for EDELWEISS-II. 
%Preliminary results are expected in 2006; the experiment could ultimately deploy up to 40 kg of detectors. Goals are 
%to gain two orders of magnitude in sensitivity and to serve as a test-bed for an even larger, ton-scale, experiment.
The EDELWEISS experiment is dedicated to the WIMPs direct search using heat-and-ionization Ge cryogenic detectors. We present the final results 
obtained by the first stage of the experiment, EDELWEISS-I which used three
320~g bolometers, corresponding to 62~kg.d. We describe EDELWEISS-II which commissioning runs have already started. This second stage of the
experiment involves 10 to 40 kg of detectors with a better shielding in the aim to improve the sensitivity by two orders of magnitude. 
\vspace{1pc}
\end{abstract}

\maketitle

\section{Introduction}
Recent cosmological observations of the CMB show that the main part of the 
matter in our Universe is dark and non baryonic~\cite{WMAP}. If non baryonic Dark Matter is made of particles,
they must be stable, neutral and massive : WIMPs (Weakly Interactive Masssive Particles). 
In the MSSM (Minimal Supersymmetric Standard Model) framework, the WIMP could be the LSP 
(Lightest Supersymmetric Particle) called neutralino. It has a mass between few tens and 
few hundreds of GeV/c$^2$, and a scattering cross section with a nucleon below $10^{-6}$~pb.\\  
The EDELWEISS experiment is dedicated to the direct detection of WIMPs. 
The direct detection principle (used also by other experiments like DAMA~\cite{dama}, CDMS~\cite{cdms1} and CRESST~\cite{cresst})
consists in the measurement of the energy released by nuclear recoils produced in an ordinary matter target 
by the elastic collision of a WIMP from the galactic halo.

\section{EDELWEISS-I}
\subsection{Experimental set-up}
The EDELWEISS experiment is located in the Modane Underground Laboratory (LSM) in the Fr\'ejus tunnel 
connecting France and Italy under $\sim$1800~m of rock ($\sim$4800~mwe). 
In the laboratory, the muon flux is 4~$\mu$/m$^2$/d and the fast neutron flux has been 
measured to be $\sim$~1.6$\times$ 10$^{-6}$~cm$^2$/s~\cite{neutron}.\\
The detectors used in the experiment are cryogenic bolometers with simultaneous measurement of
phonon and ionization signals. They are made of a cylindrical Ge crystal with Al 
electrodes to collect ionization signals and a NTD heat sensor glued onto one electrode
to collect the phonon signal~\cite{edel3}. The top electrode is segmented in a central electrode and 
an annular guard ring to define a fiducial volume corresponding to 
57~$\pm$~2~$\%$ of the total volume~\cite{edel4}. On four of the five detectors used in 
EDELWEISS-I an amorphous layer (either of Ge or Si) was deposited under the electrodes to improve 
charge collection of near surface events~\cite{surface}. \\
The detectors were operated in a dilution cryostat with a regulated temperature of 17~$\pm$~0.01~mK. 
The cryostat could not house more than 3$\times$320~g Ge detectors. 
A passive shielding made of paraffin (30~cm), 
lead~(15cm) and copper~(10~cm) surrounded the experiment~\cite{{edel1},{edel2},{lsm}}.\\

The simultaneous measurement of both heat and ionization signals provides an excellent 
event by event discrimination between nuclear recoils (induced by WIMP or neutron 
scattering) and electron recoils  (induced by $\beta$ or $\gamma$-radioactivity). The ratio
of the ionization and heat signals depends on the recoiling particle, since a nucleus
produces less ionization in a crystal than an electron does.\\
The heat and ionization responses to gamma rays were calibrated using $^{57}$Co and 
$^{137}$Cs radioactive sources and the response to nuclear recoils was measured using a 
$^{252}$Cf source. A summary of the bolometer baselines, resolutions 
and energy thresholds can be found in~\cite{edel6}.
Typically with the EDELWEISS detectors it is possible
to reject more than 99.9~$\%$ of electron recoils down to 15~keV.
%The rejection efficiency is a fundamental parameter for these types of detectors. 
%It is regularly controlled by measuring the ratio of ionization to recoil energy during
%the gamma rays calibrations, where in addition, the charge collection quality for gamma 
%rays can be checked.

\subsection{Final results}
\begin{figure}[htb]
\vspace{9pt}
\includegraphics*[width=6cm,height=6cm]{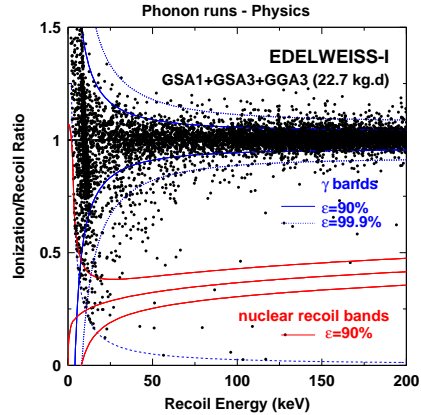}
\caption{Distribution of the ratio of the ionization energy to the recoil energy as a function of the recoil 
energy collected in the fiducial volume of three EDELWEISS detectors in the last run for 2003~\cite{edel6}.}
\label{fig:data}
\end{figure}

\begin{figure}[htb]
\vspace{9pt}
\includegraphics*[width=6cm,height=6cm]{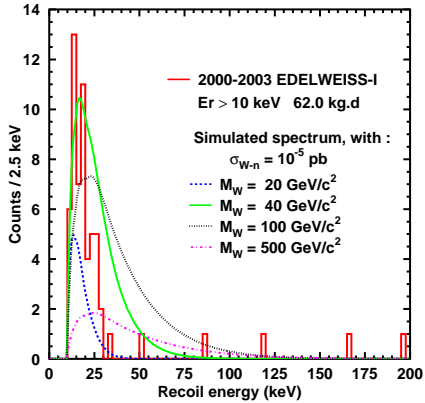}
\caption{Energy spectrum for the EDELWEISS-I data, for $E_R$~$>$~10~keV, compared to simulated 
theoretical WIMP spectrum  for M$_{W}$ = 20, 40, 500~GeV/c$^2$~\cite{edel6}.}
\label{fig:spectrum}
\end{figure}

\begin{figure}[htb]
\vspace{9pt}
\includegraphics*[width=6cm,height=6cm]{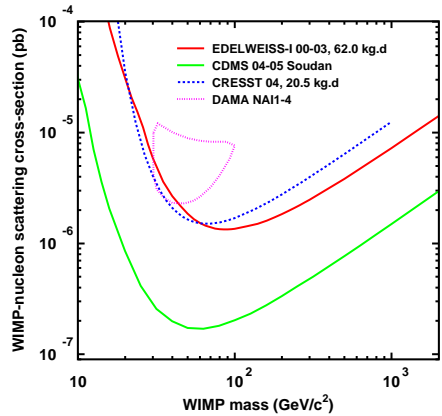}
\caption{Spin-independent exclusion limits for EDELWEISS-I without 
background subtraction~\cite{edel6} compared to the CDMS limits~\cite{cdms2}, and the CRESST 
 limits~\cite{cresst}. The closed contour is the allowed region at 3$\sigma$ C.L. from 
 DAMA NaI1-4 annual modulation data~\cite{dama1}.}
\label{fig:exclusion}
\end{figure}

During the EDELWEISS-I stage (2000-2003), four physics runs have been performed with five 
detectors. In the three first runs, the trigger was the fast ionization signal. For the
last run, the trigger was the phonon signal. Thanks to a better resolution and the absence of 
quenching factor on the phonon signal, the phonon trigger improves the efficiency 
at low energy for nuclear recoils. 
%In this last configuration, a $\sim$90$\%$ efficiency 
%has been reached at 15~keV on the three detectors, this is shown on the Figure~\ref{fig:ef}. 
%\begin{figure}[htb]
%\vspace{9pt}
%\includegraphics*[width=6cm,height=6cm]{efficacites.eps}
%\caption{.}
%\label{fig:ef}
%\end{figure}
%With the ionization trigger a $\sim$90$\%$ 
%efficiency was reached at 20~keV or 30~keV, depending on the detector. 
The overall trigger efficiency at 15 keV for the entire EDELWEISS-I data set is 50\%~\cite{edel6} 
for events in the fiducial volume.\\
The low-background physics data recorded in the phonon trigger configuration are shown 
in Figure~\ref{fig:data}.
Considering the entire 62~kg.d data set, 60 events compatible with nuclear recoils have 
been recorded above a recoil energy of 10~keV. Only 3 events have an energy between 30 
and 100 keV and this provides a strong constraint on a possible WIMP rate. 
The corresponding 
energy spectrum is shown in Figure~\ref{fig:spectrum}, compared with simulations of theoretical 
spectrum for different WIMP masses, taking into account the recoil energy dependence 
efficiency\footnote{The efficiency calculation takes into account thresholds, resolutions, the 
90~$\%$ efficiency (1.645~$\sigma$) for the nuclear recoil selection and the 99.9~$\%$ rejection 
(3.29~$\sigma$) of the electronic recoils.} of all experimental configurations. 
The 
overall shape of the experimental spectrum is incompatible with WIMP masses above 20~GeV/c$^2$.
Considering all the events with $E_R >$~15~keV 
as possible WIMP interactions and taking into account the efficiency versus recoil
energy function of each run, a conservative upper limit 
on the WIMP-nucleon cross-section as a function of the WIMP mass has been derived with the 
Optimum Interval Method~\cite{yellin}. This method allows to compute an exclusion limit in the 
presence of an unknown background without any subtraction. Figure~\ref{fig:exclusion} shows the 90 \% C.L. 
EDELWEISS-I spin-independent exclusion limits compared with that of other running experiments. 
Because of the observed counts, the new limit is consistent with the previous one, obtained with a smaller exposure~\cite{edel5}.
The best sensitivity for EDELWEISS-I is 1.5$\times$10$^{-6}$ pb at 80 GeV/c$^2$~\cite{edel6}. 
To explore more interesting supersymmetric models, a gain in sensitivity of a factor 100 is needed. This is the goal of the 
EDELWEISS-II experiment. 
This gain in sensitivity depends on improvements on the background reduction and rejection and on the increase of the detector mass. 
\\
The EDELWEISS data can also put constraints on models where spin-dependent interactions dominate~\cite{edelsd}. This is because natural Ge 
is made of 7.8 \% of $^{73}$Ge with a spin of 9/2. 
\begin{figure}[htb]
\vspace{9pt}
\includegraphics*[width=6cm,height=6cm]{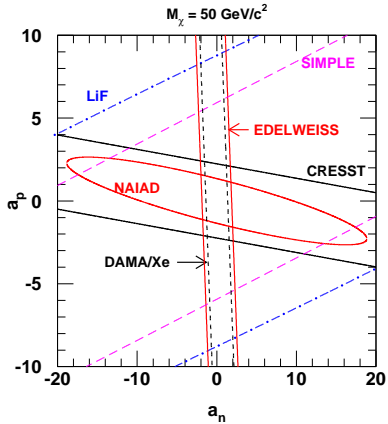}
\caption{Allowed regions in the ($a_p$,$a_n$) plane for EDELWEISS-I and other running experiments for $M_W$~=~50~GeV/c$^2$~\cite{edelsd}.}
\label{fig:spindep}
\end{figure}
Figure~\ref{fig:spindep} shows the constraints established for different experiments in the ($a_p$,$a_n$) plane, where $a_{p,n}$ are the
effective WIMP-proton (-neutron) couplings. 
Experiments with Ge detectors are the most sensitive to $a_n$, but optimistic models are still a factor 100 below the present 
limits~\cite{edelsd}.

\subsection{Background studies}
Although the EDELWEISS-I limits of Figure~\ref{fig:exclusion} are derived assuming all events as 
possible WIMP candidates, the experimental data reveals some clues as to the nature of 
possible backgrounds. \\
From the simulation of the neutron flux in the laboratory~\cite{neutron}, we expect 2 nuclear recoils in 
62 kg.d with $\sim$~10~$\%$ being coincidences. One two-detector coincidence of nuclear recoils has been 
recorded. This event is very likely a neutron-neutron 
coincidence, indicating that a certain fraction of events in Figure~\ref{fig:spectrum} could be due
to single hits by neutrons. 
The measurement is statistically consistent with the prediction but does not provide a strong constraint on the single rate.\\
Miscollected charge events, as indicated by the few events lying 
between electronic and nuclear recoil bands in Figure~\ref{fig:data}, are another possible source of 
background But with the present statistics, 
largely limited by the number of detectors, it is not possible to conclude any further. \\
The gamma ray background is well understood. It results from the copper shielding present 
around the cryostat.  Most of the rate observed at high and intermediate energy can be explained by the measured contamination in U and 
Th of our copper shielding which will be removed for the future experiments.\\  
The contributions of $\alpha$-emitters and their daughters was investigated by studying high energy events.  
A peak appeared at a recoil energy of 5.3 MeV with a quenching factor of 0.3. This is attributed to alpha decays from $^{210}$Po near the detector
surface. 
It seems very likely that these events are due to $^{210}$Pb contamination on the surface of Cu facing detectors, because no peak at 100 keV 
from the lead recoils has been observed~\cite{simon}. By removing the Cu cover, these events should disappear.

\section{EDELWEISS-II} 
EDELWEISS-II is the second stage of the experiment which installation is achieved in november 2005. 
The first phase consists in 28 detectors with NTD or NbSi sensors. 
The expected sensitivity of this second phase is 0.002~evt/kg/d. 
Specific improvements are aimed at reducing the possible background sources, 
that may have limited the sensitivity of EDELWEISS-I.\\
To reduce the radioactive background in the cryostat all the materials are tested for 
radiopurity in a HPGe dedicated detector, with very low radon level. A clean room surrounds the experiment 
and air flow is available.\\
Concerning the low energy neutron background, due to the radioactive surrounding rock, it is attenuated 
by more than three orders of magnitude thanks to a 50~cm polyethylene shielding. 
In addition, a muon veto~\cite{veto} surrounding the experiment will tag muons interacting
in the lead shielding. The increased number of detectors, up to $\sim$110 in a compact arrangement, will improve the possibility of 
detecting multiple interactions of neutrons.\\
One important limitation for the EDELWEISS-I sensitivity is the presence of surface events, namely interactions near electrodes. 
Because of diffusion, trapping and recombination, the charge in surface events is miscollected and can mimic nuclear recoils.
Two solutions exist. First, the passive rejection in improving the charge collection of surface events with amorphous layer or with 
thick electrodes. And second, the active rejection in identifying the surface events with the pulse shape analysis or NbSi sensors. \\
One of the R$\&$D goals in EDELWEISS is the event-by-event identification of these miscollected 
events and their active rejection. 
A new generation of detectors has been developped with NbSi thin film sensors (instead 
of the NTD heat sensors for present detectors). They consist in a Ge crystal with 
two NbSi sensors acting also as electrodes for charge collection. These thin film sensors 
are sensitive to the athermal component of the phonon signal, acting as near-surface 
interaction tag~\cite{nbsi}. Several tests have been made in the EDELWEISS-I setup with 
three 200~g Ge detectors showing an improvement of a factor 20 of the rejection and only a reduction 
of 10\% of the fiducial volume. With these very encouraging results, seven 400~g Ge detectors are being prepared 
for the first stage of EDELWEISS-II.\\
In January 2006 a first cryogenic run has been recorded. Presently, the experiment is in commissioning run in different configurations 
with 8 bolometers to valid a lot of items : cryogeny, detectors, microphony, new electronic and acquisition system, ...

\section{Conclusion}
EDELWEISS-I experiment has reached its limit sensitivity near 10$^{-6}$~pb, allowing 
the exclusion of some optimistic SUSY models. The presence of two possible backgrounds have been observed, neutrons and surface events. 
The goal for the future with EDELWEISS-II is to gain a factor 100 in sensitivity thanks to an improved setup.
EDELWEISS is pursuing its R\&D on the detectors with NbSi or NTD sensors to permit the identification of surface events.
EDELWEISS-II will prepare the next generation of detector in the 100 kg to one ton scale with the EUECA project~\cite{eureca}.

\end{document}